\documentclass[universe,article,accept,pdftex,moreauthors]{Definitions/mdpi}
\firstpage{1} 
\pubvolume{11}
\issuenum{1}
\articlenumber{0}
\pubyear{2025}
\copyrightyear{2025}
\externaleditor{Stefano Vercellone} 
\datereceived{12 August 2025} 
\daterevised{14 October 2025} 
\dateaccepted{4 November 2025} 
\datepublished{ } 
\hreflink{https://doi.org/} 

\Title{A New Phase of Optical Activity of BL Lacertae in the Fall of 2024: Intra-Night Flux and Polarization Variations}
\TitleCitation{A New Phase of Optical Activity of BL Lacertae in the Fall of 2024: Intra-Night Flux and Polarization Variations}

\Author{{Rumen Bachev
} $^{1,}${*}\orcidA{}, Milen Minev $^{1}$, Anton Strigachev $^{1}$ and Alexander Kurtenkov $^{1,2}$}

\AuthorNames{R. Bachev, M. Minev, A. Strigachev and A. Kurtenkov}

\AuthorCitation{{Bachev,} R.;  Minev, M.; Strigachev, A.; Kurtenkov, A.}

\address{%
$^{1}$ \quad Institute of Astronomy and NAO, Bulgarian Academy of Sciences, 72 Tsarigradsko Shose, 1784 Sofia, Bulgaria; mminev@astro.bas.bg (M.M.); anton@nao-rozhen.org (A.S.); al.kurtenkov@abv.bg (A.K.)\\

$^{2}$ \quad Faculty of Physics, Sofia University ``St. Kliment Ohridski'', 5 James Bourchier Blvd., 1164 Sofia, Bulgaria}

\corres{Correspondence: bachevr@astro.bas.bg}

\abstract{BL Lacertae is not only archetypical of an entire class of jet-dominated active galactic nuclei, blazars, but also one of the most active and rapidly changing objects in this class. In the fall of 2024 (September--November), BL Lacertae underwent another episode of strong optical activity, reaching an R-band magnitude of about 12 and showing extremely rapid and large-amplitude inter- and intra-night flux and polarization variations. During this period, the object was monitored over 40 nights using telescopes with an aperture of up to 2 m at three observatories: Rozhen and Belogradchik in Bulgaria and Skinakas in Greece. The results from this study include some of the most spectacular intra-night variability episodes detected in a blazar. These rapid variations, combined with high photometric accuracy and high time resolution, allowed for confirmation of consistency between different optical bands with zero time delays, down to a minute scale. Unlike previous activity reports, polarization was relatively stable on these short time-scales. Possible connections between polarization, flux, and intra-night variability were explored in order to better model or constrain the physical processes and emission mechanisms in the relativistic jets.}


\begin{document}

\section{Introduction}
BL Lacertae ($\alpha$ = 22~02~43.3
, $\delta$ = +42~16~39, $z=0.066$), which lent its name to an entire class of active galactic nuclei (AGNs)---blazars---was initially considered a variable star of irregular type and was correspondingly named as such. Blazars, which are crucial to understanding high-energy physics and AGN classes, are among the most studied extragalactic objects. Blazars are jet-dominated AGNs, in which most emissions are produced in a relativistic jet via non-thermal processes (synchrotron or Compton scattering, particle cascades, etc.). To be classified as a blazar, the object's relativistic jet has to be oriented at a small angle from the observer’s line of sight, making the entire blazar emission significantly Doppler-{boosted} 
 \citep{Urry95}.

Due to the nature of their emission mechanisms and significant Doppler boosting, blazars exhibit very unique observational properties atypical for the rest of the AGN population. Among these properties are an extremely broad, two-humped spectral energy distribution (SED), often stretching radio to UHE $\gamma$-rays; significant polarization; large and violent variations across all emissions; apparent superluminal motions; and, as discovered recently, possible neutrino production \citep{IC18a, IC18b}. However, not all objects manifest all these distinctive properties all of the time. This is primarily because the blazar class can be divided into subclasseses, and the objects belonging to different subclasses may exhibit different properties. Blazars are often divided into subclasses based on the location of the two humps in their SEDs. The low-energy hump/peak is most likely a natural synchrotron, but the origin of the high-energy hump/peak is still under debate. One proposed explanation is based on leptonic models, in which Compton scattering in the form of a synchrotron-self Compton (SSC) or external Compton (EC), depending on the origin of the seed photons, is responsible for generating the high-energy emissions. Alternatively, in hadronic models \citep{Bott13}, proton synchrotron emission and/or particle cascades are thought to produce the high-energy peak.

For low-synchrotron-peaked (LSPs) objects, the synchrotron emission peaks at $\nu_{\rm syn}^{\rm peak} \leq 10^{14}$ Hz, while for the high-synchrotron-peaked (HSPs) objects, it peaks at $\nu_{\rm syn}^{\rm peak} \geq 10^{15}$ Hz \citep{Abdo10}. Those with peaks in between are classed as intermediate (ISP) types. Among LSP objects, some show additional thermal features in their low-energy spectra, indicating significant contributions of a thin accretion disk and a broad line region; these objects are also classified as flat-spectrum radio quasars (FSRQs). There is emerging evidence suggesting that LSP objects are more variable and show higher polarization in the optical region (where the synchrotron peak is located for these objects), while HSP objects show low-level activity in this region at best. This is especially true for intra-night time-scale variability; over the years, practically only LSP objects (or FSRQs) have been reported to show violent flux changes on such short time-scales \citep{Hov14, Zha15, Bach18, Anj20, Zha24}. On the other hand, HSP objects appear to be more active in higher energy regions (X- and $\gamma$-rays). As a rule, violent activity is not present all the time, even for the most active objects, i.e., there are relatively active and quiet periods for every object. The highest activity episodes, sometimes including quasi-periodic oscillations, are often associated with the highest flux states \citep{Jors22}.

BL Lacertae is typically classified as an LSP/ISP (non-FSRQ) object \citep{Wand25} and is among the most active blazars with respect to variability. The object has shown extremely variablility in both flux and polarization during some periods of activity, one of the most recent of which occurred in 2020--2021 \citep{Imaz23, Shab23, Bach23}. During that period, in addition to unprecedented high optical flux levels (reaching $R\simeq11$ mag), BL Lacertae often showed optical flux changes of more than 0.1 mag for a half an hour and polarization changes as fast as 1--2\% in polarization degree and 5--10$^{\circ}$ in polarization angle for an hour \citep{Bach23}.

Later optical and high-energy observations, however, suggest that BL Lacertae entered another activity period in the fall of 2024. Starting in September 2024, the object showed significantly enhanced long-term $\gamma$-ray (>100 MeV) activity, which was detected using \textit{Fermi
}-LAT (including a historical maximum in October 2024 \citep{vZyl24}), as well as activity at very high energies, detected with ground-based facilities such as LHAASO, MAGIC, and VERITAS \citep{Xia24, Pan24, Furn24}. This high activity state was also confirmed in the X-ray, UV, and optical bands by \textit{Swift} \citep{Noz24} and ground-based optical observations \citep{Kish24}. In addition, BL Lacertae significantly increased its intra-night optical variability \citep{Bach24a}. In this paper, we study this optical activity by presenting and discussing the results of extensive multi-color flux and polarization monitoring of BL Lacertae on intra-night time-scales. {So far, violent intra-night variability episodes have been detected in the optics of several blazars (virtually all of which are classified as LSP/FSRQ objects) and reported, including AO 0235+164 \citep{Rome00}, S5 0716+714 \citep{Bhat16}, PKS 0736+017 \citep{Clem03}, S4 0954+65 \citep{Bach15, Rai23b}, BL Lacertae \citep{Rai23}, CTA 102 \citep{Bach17}, and 3C 454.3 \citep{Bach11}.}

\section{Observational Data}

Between May 2024, and February 2025, BL Lacertae was observed over 41 nights with three telescopes: the 2 m RCC telescope of Rozhen National Observatory, the 60 cm telescope of Belogradchik Observatory, both located in Bulgaria, and the 1.3 m telescope of Skinakas Observatory, Crete, Greece. All telescopes were equipped with cooled CCD cameras and standard filter sets. In addition, the Belogradchik telescope was equipped with polarizing filters to measure the linear polarization in selected optical bands (details are provided in \citep{Bach23,Bach24b}). Multicolor intra-night monitoring was performed over the course of 31 nights for a total of about 125 h (about 4 h per night on average). The multicolor monitoring was usually quasi-simultaneous, i.e., using repeating (B)VRI filters with typical exposure times of 60--120 s. The BVRI sequences were interrupted repeatedly to measure the polarimetry (Belogradchik telescope), which was measured in the R-band. {Typical photometric errors for the Belogradchik 60-cm telescope are in the range of 0.01--0.02 mag for VRI bands and about 0.02--0.04 mag for B-bands. The Skinakas 1.3 m telescope was used for monitoring on four nights and has a typical photometric uncertainty of about 0.01 mag.} The 2 m Rozhen telescope was used (five nights) for \textit{simultaneous} monitoring in two optical regions, using a 570 nm beam splitter and two different \textit{Andor} cameras (in focal reducer mode). Only the beam splitter was used, with no additional filters, to improve the photometric accuracy and time resolution of the light curves. The lack of a broadband filter should not greatly influence the results, since the primary goal wa to study intra-night variability characteristics, including possible wave-dependent time delays. This approach enabled the exposure time to occasionally be reduced to 4 s (typically, 7--20 s is required for the red channel and 30--60 s is required for the blue channel), with photometric accuracy of {0.003--0.005 mag (red channel) and 0.005--0.01 mag (blue channel)}. This allowed for the capture of one of the best examples of violent blazar variability ever reported, captured simultaneously in different optical regions (see the next section). {The measured photometric uncertainties were further substantiated by the very similar standard deviations of light curves of stars, which were measured against each other.}

All frames were flat-field and dark/bias-corrected. Standard aperture photometry was performed with an aperture radius of 8 arcsec. {This aperture is used for this object in all WEBT/GASP campaigns 
 (\url{https://www.oato.inaf.it/blazars/webt/introduction/}, accessed on 11 November 2025)}. {The choice of this aperture radius is justified in \citep{Rai23} and aims to include the host galaxy of BL Lacertae entirely without including too many other faint stars in this otherwise rich field. The host galaxy's magnitude is estimated to be R $\simeq$ 15.55 (e.g., \citep{Sca00, Rai09}), significantly fainter than the blazar during high-activity periods and significantly smaller in size than the aperture radius (the half-light radius of the host galaxy is estimated at 4.8 arcsec \citep{Sca00}). Visual inspection of our frames indicated the presence of one star of about 20.2 mag (R-band) within the aperture radius, whose impact on the measured variability should be considered negligible even under varying seeing conditions.} 

Nearby comparison stars (B, C, and H {from \citep{Fio96}; their positions and calibrated magnitudes are provided onn the GASP website, 
 \url{https://www.oato.inaf.it/blazars/webt/2200420-bl-lac/}, accessed on 11 November 2025}) were used for magnitude calibration. All comparison stars were found to be photometrically stable within their respective photometric uncertainties {\citep{Pac13}}. {All observation nights ranged from clear and photometric to moderately cloudy, with seeing {typically between 1 and 4 arcsec (see 
   the next section).}}

\begin{figure}[H]
\includegraphics[width=140mm]{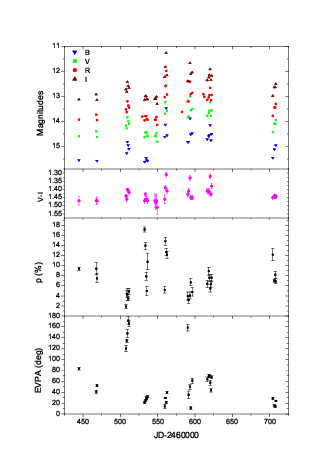}
\caption{Long-term 
 photometric and polarimetric behavior of BL Lacertae between May 2024, and February 2025. From top to bottom: Light curves in BVRI, V-I colors, polarization degrees, and polarization angles (both measured in the R-band).}
\label{f1}
\end{figure}

\begin{figure}[H]
\includegraphics[width=140mm]{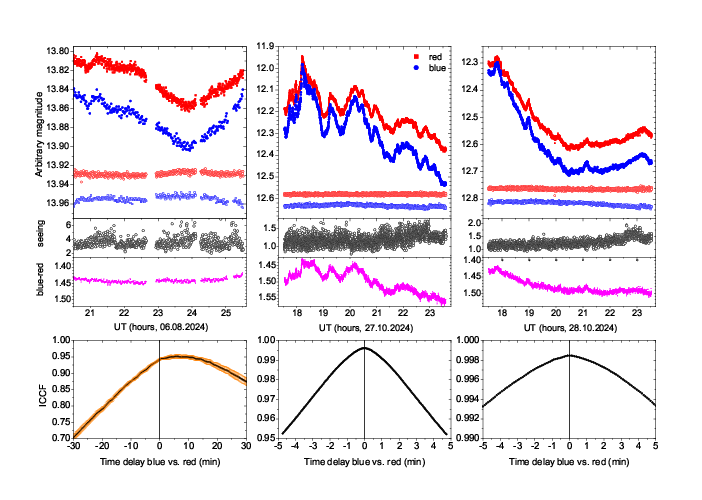}
\caption{Intra-night 
 variability in two optical regions, divided by a beam splitter at 570 nm (see the text), obtained simultaneously using a 2 m Rozhen telescope over three different nights {(indicated below the middle panels)}. {The top panel for each night displays the light curves of the blazar (upper two curves) and a comparison star, C, (lower two curves), measured with respect to the other {reference stars}. Verical shifts of the light curves were applied for presentation purposes. The {next two panels show the seeing during the night (in arcsec) and} the (arbitrary) color changes of the blazar on intra-night time-scales.} The bottom panel presents the interpolation cross-correlation as a function of time lag between the bands (a positive time peak means the blue range is leading). To assess the reality of the positive peak for the night of {6 August 2024} ({left panels}), an additional MC simulation was performed, and error bars were added (see the text). Note the extremely violent variability recorded during the other two nights, with virtually zero lag between the bands. Also note the color changes, which indicate BWB behavior on intra-night time-scales ({middle panels}).}
\label{f2}
\end{figure}


\begin{figure}[H]
\includegraphics[width=140mm]{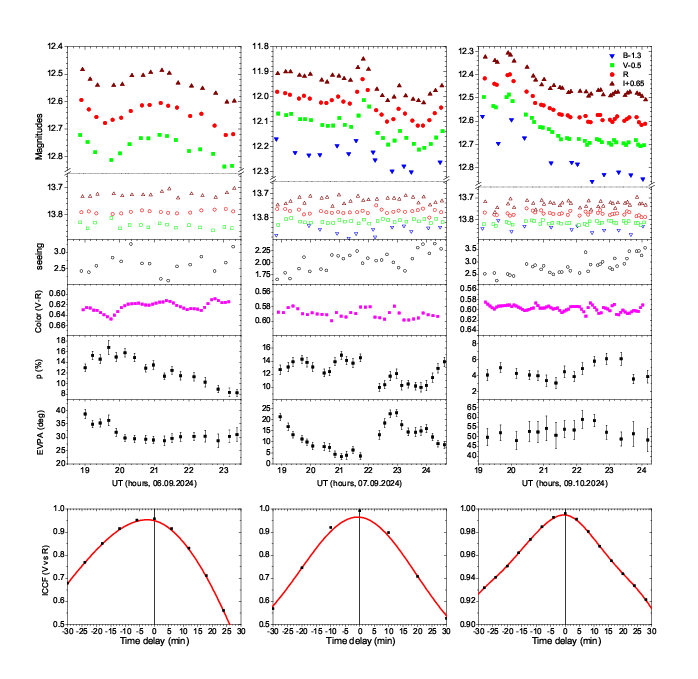}
\caption{Quasi-simultaneous 
 intra-night photometric and polarimetric monitoring of BL Lacertae for 3 nights {(indicated above the bottom panels)}, performed with the 0.6 m Belogradchik telescope. {The top panel for each night presents the light curves of the blazar and a comparison star (C measured with respect to B; see the text). Arbitrary magnitude shifts were applied to each light curve for clarity, and these shifts are indicated for BL Lacertae. The next {4} panels display the {seeing variations during the night (in arcsec), the} color changes, PD, and PA (measured in the R-band), from top to bottom.} The lower panel shows the interpolation cross-correlation lags between the V and R bands (positive peaks indicate a leading V-band). Splines (in red) are provided to guide the eye. Linear interpolation was also used to plot the V-R color, where the small scale structures are likely an artefact of this interpolation. The typical photometric errors of the light curves are comparable to the symbol size (see the text).}
\label{f3a}
\end{figure}

\begin{figure}[H]
\includegraphics[width=140mm]{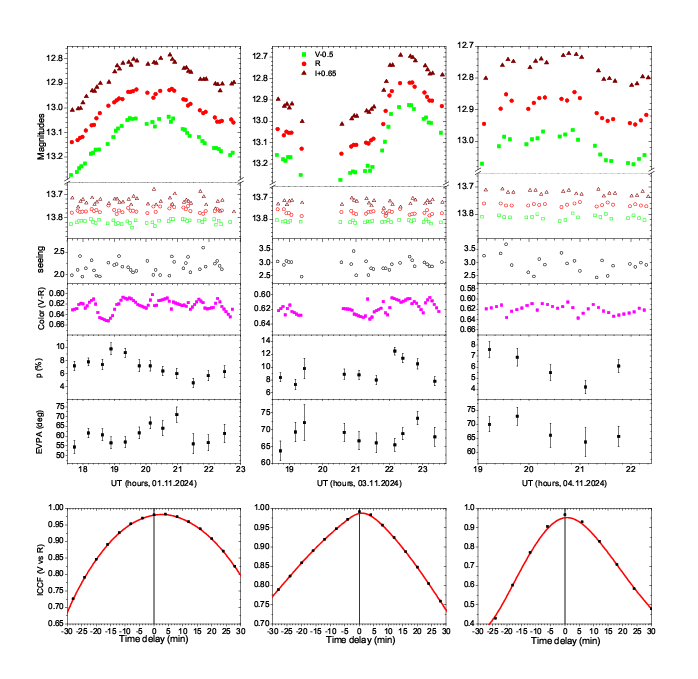}
\caption{See 
 Figure \ref{f3a}.}
\label{f3b}
\end{figure}

\begin{figure}[H]
\includegraphics[width=140mm]{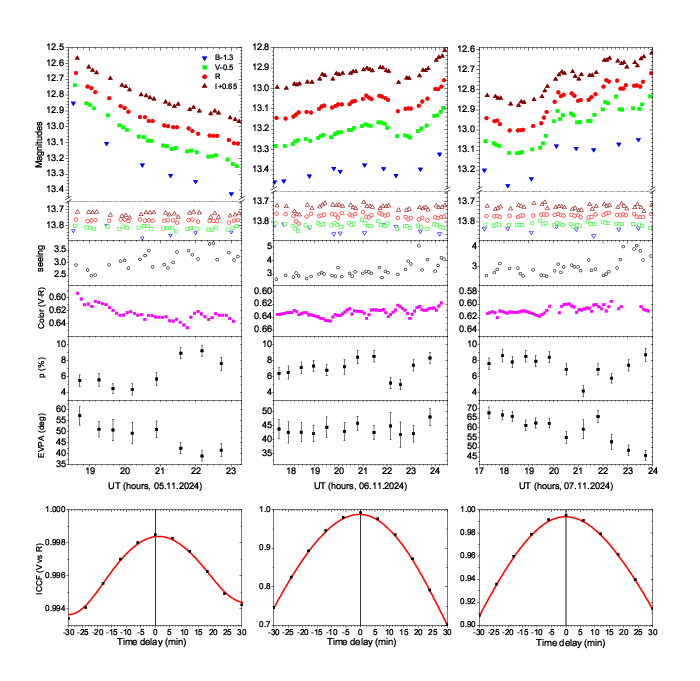}
\caption{See 
 Figure \ref{f3a}.}
\label{f3c}
\end{figure}


\section{Results}

Table \ref{tab:log} presents a log of the observations, with some of the most important measurements, including the average R-magnitudes, colors, PD, and PA (in percents and degrees, respectively), as well as the duration of monitoring (in hours) if applicable. {No attempts were made to correct any of the measured quantities for the presence of the host elliptical.} {A letter after the evening date indicates the stability of that night during the intra-night monitoring epochs: s (stable---clear and photometric), ps (partly stable---cirrus clouds), and u (unstable---passing clouds other weather-related problems).} These data are also presented in Figure \ref{f1}, where BL Lacertae's long-term (May 2024--February 2025) variability in four colors is displayed (Figure \ref{f1}, upper panel). The new activity period started in the first days of September (2024), when the brightness reached $R\simeq12$ mag (7 September), in contrast to earlier typical values of 13--14 mag. This high optical state was accompanied by rapid brightness and polarization changes on intra-night time-scales (Figures \ref{f2}--\ref{f3c}). Linear polarization, on the contrary, did not seem to be significantly affected by this outburst in terms of polarization degree (PD) or polarization angle (PA or EVPA). The polarization degree was typically around  10\% (between 2 and 20\%), while the PA covered almost the entire realm of possibilities (0--180$^{\circ}$), with some concentration around 20$^{\circ}$ (Figure \ref{f1}, lower two panels). These polarization results are similar to those reported during the previous activity state \citep{Rai23}. As such, they cannot be related to the brightness in any way \citep{Bach23}. A slight bluer-when-brighter behavior can also be traced (Figure \ref{f1}, second panel from the top), which is typical for this and other BL Lac-type objects 
 \citep{Meng17, Li21, Fang22}.

\begin{table}[H]
\caption{Log of observations.} 
\label{tab:log} 



\begin{adjustwidth}{-\extralength}{0cm}

\newcolumntype{C}{>{\centering\arraybackslash}X}
\begin{tabularx}{\fulllength}{crCCCCCCC}

\toprule
\textbf{JD}	&	\textbf{Evening}	&	\textbf{Telescope}	&	\textbf{<R>}	&	\textbf{<V-I>}	&	\textbf{<PD>}	&	\textbf{<PA>} 	&	\textbf{Duration} 	&	\textbf{rms}	\\

\midrule															
2,460,444.567 
	&	13.05.24	&	B60	&	13.93	&	1.47	&	9.4	&	82.9	&		&		\\
2,460,467.537	&	05.06.24	&	B60	&	13.73	&	1.47	&	9.4	&	40.7	&		&		\\
2,460,468.527	&	06.06.24	&	B60	&	13.95	&	1.47	&	7.5	&	52.1	&		&		\\
2,460,507.478	&	15.07.24 s	&	B60	&	13.52	&	1.44	&	1.9	&	120	&	2.5	&	0.019	\\
2,460,508.469	&	16.07.24 ps	&	B60	&	13.63	&	1.46	&	4.4	&	134.3	&	2.5	&	0.009	\\
2,460,509.471	&	17.07.24 ps	&	B60	&	13.20	&	1.40	&	4.3	&	148	&	2	&	0.052	\\
2,460,510.455	&	18.07.24 u	&	B60	&	13.37	&	1.41	&	3.5	&	171.6	&	3	&	0.021	\\
2,460,511.413	&	19.07.24	ps &	B60	&	13.45	&	1.42	&	4.9	&	166.1	&	2	&	0.016	\\
2,460,529.314	&	06.08.24	u&	R200	&	13.81 $^{a}$ 
	&		&		&		&	5.1	&	0.016  $^{b}$	\\
2,460,530.292	&	07.08.24	ps&	R200	&		&		&		&		&	6.5	&	0.028 $^{c}$	\\
2,460,532.490	&	09.08.24	u&	B60	&	13.96	&	1.47	&	17.2	&	20.9	&	4.5	&	0.018	\\
2,460,533.492	&	10.08.24 s	&	B60	&	13.79	&	1.43	&	14	&	24.6	&		&		\\
2,460,534.446	&	11.08.24	s&	B60	&	13.79	&	1.46	&	7.9	&	29.2	&	3	&	0.018	\\
2,460,535.413	&	12.08.24	s&	B60	&	13.91	&	1.46	&	5	&	28.8	&	4.5	&	0.015	\\
2,460,536.332	&	13.08.24	s&	B60	&	13.93	&	1.47	&	10.8	&	31.6	&		&		\\
2,460,546.308	&	23.08.24	s&	Sk130	&	13.93	&	1.47	&		&		&	2.4	&	0.01	\\
2,460,547.299	&	24.08.24	s&	Sk130	&	13.94	&	1.48	&		&		&	2.4	&	0	\\
2,460,548.297	&	25.08.24	s&	Sk130	&	13.83	&	1.47	&		&		&	2.4	&	0.02	\\
2,460,549.293	&	26.08.24	s&	Sk130	&	14.14	&	1.51	&		&		&	2.4	&	0.01	\\
2,460,559.338	&	05.09.24 ps	&	B60	&	13.03	&	1.46	&	5.2	&	13.9	&		&		\\
2,460,560.287	&	06.09.24 ps&	B60	&	12.59	&	1.39	&	14.9	&	29.4	&	4.4	&	0.03	\\
2,460,561.286	&	07.09.24 s	&	B60	&	11.98	&	1.31	&	12.7	&	21.3	&	6	&	0.04	\\
2,460,562.278	&	08.09.24 s	&	B60	&	12.93	&	1.41	&	12.2	&	39.2	&	5	&	0.01	\\
2,460,582.407	&	28.09.24	&	R200	&	13.62 $^{a}$	&		&		&		&	2	&	0.019 $^{b}$	\\
2,460,590.375	&	06.10.24 u&	B60	&	13.20	&	1.43	&	4	&	158.7	&	3	&	0.05	\\
2,460,591.251	&	07.10.24 s&	B60	&	13.20	&	1.41	&	3.2	&	35.1	&	6.4	&	0.08	\\
2,460,593.300	&	09.10.24 s&	B60	&	12.42	&	1.33	&	4.1	&	49.8	&	4.9	&	0.07	\\
2,460,594.322	&	10.10.24 ps	&	B60	&	12.91	&	1.45	&	6.7	&	11	&	2.6	&	0.04	\\
2,460,596.271	&	12.10.24 s	&	B60	&	12.86	&	1.45	&	4.8	&	61.2	&	2.5	&	0.03	\\
2,460,611.231	&	27.10.24 s	&	R200	&	12.91 $^{a}$	&		&		&		&	6	&	0.09 $^{b}$	\\
2,460,612.229	&	28.10.24 s	&	R200	&	13.02 $^{a}$	&		&		&		&	6	&	0.1 $^{b}$	\\
2,460,616.235	&	01.11.24	ps&	B60	&	13.14	&	1.41	&	6.4	&	64.1	&	5.2	&	0.06	\\
2,460,618.277	&	03.11.24	s&	B60	&	13.04	&	1.41	&	8.9	&	69.2	&	4.9	&	0.11	\\
2,460,619.296	&	04.11.24	s&	B60	&	12.95	&	1.42	&	7.6	&	70	&	3.2	&	0.03	\\
2,460,620.276	&	05.11.24	s&	B60	&	12.66	&	1.32	&	5.5	&	57.2	&	4.6	&	0.12	\\
2,460,621.225	&	06.11.24	s&	B60	&	13.15	&	1.43	&	6.4	&	43.8	&	7	&	0.1	\\
2,460,622.220	&	07.11.24	s&	B60	&	12.94	&	1.38	&	7.6	&	67.7	&	6.6	&	0.08	\\
2,460,704.208	&	28.01.25	&	B60	&	13.78	&	1.45	&	12.2	&	28.5	&		&		\\
2,460,706.209	&	30.01.25	&	B60	&	13.44	&	1.44	&	7	&	15.8	&		&		\\
2,460,707.259	&	31.01.25	&	B60	&	13.45	&	1.45	&	8.2	&	13.9	&		&		\\
2,460,708.210	&	01.02.25	&	B60	&	13.30	&	1.44	&	6.9	&	23.9	&		&		\\
\bottomrule
\end{tabularx}
\end{adjustwidth}

\noindent{\footnotesize{$^{a}$ The R-band magnitude is approximate, as no real R-filter was used (see the text).
 $^{b}$ Red range ($>$570 nm) variations were used to calculate the rms (see the text).
 $^{c}$ Blue range ($<$570 nm) variations were used to calculate the rms (see the text).
Telescope 
 codes:
B60: 60-cm Cassegrain telescope, Belogradchik Astronomical Observatory, Bulgaria.
R200: 200-cm RCC telescope, Rozhen National Astronomical Observatory, Bulgaria. 
Sk130: 130-cm Skinakas telescope, Skinakas Observatory, Crete, Greece.
Weather
 codes (see the text):} s---stable; ps---partly stable; u---unstable.}

\end{table}

\vspace{-6pt}
Figure \ref{f2} shows three remarkable examples of extreme intra-night variations in two wavebands, which are among the most spectacular such events ever documented (especially considering the nights of 27 and 28 October 2024). These data were obtained with the 2 m Rozhen telescope, as described in the previous section. All magnitudes are arbitrary, but the red channel should roughly correspond to the R-band (the blue channel is shifted for clarity). Intra-night light curves (upper panels), color changes (middle panels), and cross-correlations between bands (lower panels) are shown for three nights. {Each upper panel also includes star C, measured with respect to some of the other {reference stars} in the field (the lower two curves were properly shifted for presentation purposes).} The lag-dependent cross-correlation function ({ICCF}) was computed by applying the interpolation cross-correlation method \citep{Gask86}. Note the significant change in variability before (left panels) and after the start of the new activity period (middle and right panels). The left panels (the night of 6 August 2024
) show indications of a possible time delay between bands. For this case, we also performed MC analysis to assess the errors of the cross-correlation function. Based on this analysis, we cannot reject the possibility that the delay of the red band behind the blue band may be real. This conclusion, however, is based on very low-level variations in both bands and could simply be due to a few peculiar deviations in the curves. No such delays were found during the next two nights of observation with this instrument, when the variations were much more rapid and prominent. We also noted that during the high-activity period (the nights of 27 and 28 October, Figure \ref{f2}), not only was intra-night variability extremely violent, reaching up to 0.4 mag for about 2 h (the night of 28 October), but color changes were also clearly evident (Figure \ref{f2}), roughly mimicking overall changes in brightness, which is an indication of bluer-when-brighter behavior (see the next section). Figures \ref{f3a}--\ref{f3c} show examples of BL Lacertae's intra-night flux and polarization variability, which were obtained with the 60 cm Belogradchik telescope. {Each top panel shows the light curves in the (B)VRI band of BL Lacertae (upper part) and star C, measured with respect to B (lower part). Except for the R-band, magnitude shifts were applied for presentation purposes (as indicated in the figures for the light curves of the blazar).} The cross-correlation between the bands is also shown; however, the time resolution here is much smaller due to the use of repeating filters. We see behavior very similar to that reported during the most recent activity period \citep{Bach23}, namely, rapid but almost independent variations in any of the three measurable quantities---flux, polarization degree, and polarization angle. An outburst during the night of 3 November 2024, where the object increased its brightness by 0.3 mag for about 30 min, is worth mentioning.

\section{Discussion}

\subsection{Polarization Changes}

Studying blazar polarization can help determine the origin of the high-energy peaks in blazar SEDs, at least with respect to LSP/ISP objects. This is especially true following the opening of the \textit{IXPE} space observatory \citep{Weis22}, which is dedicated to studying the soft X-ray polarization of astrophysical sources. For LSP/ISP objects, the soft X-ray region is part of the high-energy peak, so one can directly compare optical (low-energy peak) and X-ray (high-energy peak) polarizations, considering that the leptonic and hadronic models predict rather different $p_{\rm opt}/p_{\rm X}$ ratios (see Table 1 in \citet{Agu25}
). During a campaign in the fall of 2023, \citet{Agu25} reported $p_{\rm opt}/p_{\rm X}>3$, which they interpreted as evidence supporting the leptonic model. {Their arguments are based on the fact that the Compton scattering assumed by the leptonic models is expected to significantly reduce the polarization of the seed photons \citep{Zhan24}, while the hadronic models (in the form of either proton synchrotrons or particle cascades) ultimately rely on the synchrotron (polarized) emission of charged particles. On the other hand, \citet{Tave25} argued that the different spectral slopes of the two peaks at the locations of the polarization measurements can explain the lower X-ray polarization, even if hadronic processes are at work. Debate continues over which processes produce the high-energy peak.} Interestingly, during \citet{Agu25}'s campaign, BL Lacertae reached a record high optical polarization degree of $p_{\rm opt}\simeq 48$\% (November 2023).

Synchrotron radiation is naturally polarized, but the degree of polarization depends on how ordered the magnetic field is and/or how many independent emitting cells (with randomly oriented $B$) contribute to the emission. Clearly, $p_{\rm tot}\simeq p_{\rm max}/\sqrt{N}$ for $N$ contributing cells with {randomly oriented, uncorrelated} magnetic fields. The polarization variability, especially on intra-night time-scales, should be related to very fast processes, like the passage of plasma blobs emitted through standing shocks along the jet, where the magnetic field will be significantly compressed (and ordered \citep{Mars21}). We note that during this 2024 campaign, in terms of intra-night polarization variability, BL Lacertae was not as active as it was during the previous 2020 campaign. However, polarization changes are clearly present as both PD changes (e.g., on nights of 6 September 2024 and 5 November 2024) and PA changes (e.g., on the nights of 7 September 2024 and 7 November 2024), as is evident in Figures \ref{f3a} and \ref{f3c}.

\subsection{Fractional Variability}

\begin{figure}[H]
\includegraphics[width=138mm]{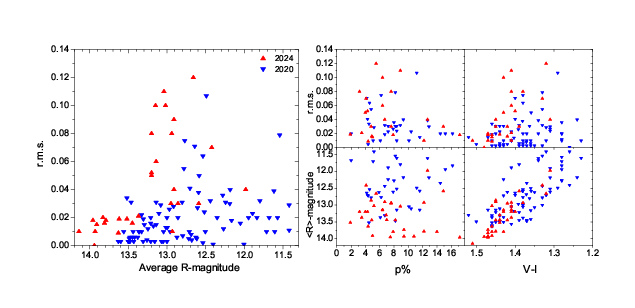}
\caption{Comparison of the long-term and intra-night flux and polarization characteristics of the 2020 and 2024 campaigns. \textbf{Left panel}: The relationship between the fractional variability (r.m.s.) and the average \textit{R}-band magnitude. A slight tendency of very little statistical significance for higher fractional variability during episodes of higher flux levels can be traced. The four \textbf{right panels} show relationships among the flux, color, r.m.s. and polarization degree. No clear dependencies are seen except for the color--magnitude relationship (BWB), which is expected for this type of blazar.}
\label{f4}
\end{figure}

Figure \ref{f4} compares some photometric, polarimetric, and variability parameters from the 2020 activity phase \citep{Bach23} and this study. We searched for possible relationships among the average flux, color, the polarization degree, and the intra-night fractional variability, $\sigma_{\rm rms}$, and compared the results from both campaigns (Figure \ref{f4}). We defined the fractional variability as $\sigma_{\rm rms} = \sqrt{\frac{1}{N-1}\sum_{i=1}^{N} (m_{i}-\langle m \rangle)^{2} - \langle \sigma_{\rm phot} \rangle^{2}}$, where $m_{i}$ are the individual magnitude measurements, $\langle m \rangle$ is the average magnitude, $N$ is the number of data points, and $\langle \sigma_{\rm phot} \rangle$ is the average photometric uncertainty. The fractional variability is considered zero for negative expressions under the square root. We note that for both campaigns, the average time duration of the intra-night monitoring datasets is almost equal (about 4 h), which eliminates any length-related issues during comparison. The most notable difference between the two activity episodes appears to be the significantly higher intra-night variability for the slightly lower average flux during the 2024 campaign. The combined data from both campaigns indicate that the polarization degree does not correlate with the average flux or the intra-night variability, which was also noted by \citet{Shko25} for this object. 

\subsection{The Shortest Variability Time-Scale}

The shortest variability time-scale, often defined as $t_{\rm var}\simeq \frac{\langle F \rangle }{|dF/dt|} \simeq \frac{1}{|dm/dt|}$, {where $dm/dt$ is the magnitude change rate}, is sometimes associated with the central black hole parameters. However, considering that the emission is produced very far from the center, it is more relevant to use this value to restrain the size of the emitting region or the electron cooling time. If the shortest variations, for instance, are due to synchrotron losses, then $t_{\rm cool} \leq \frac{\delta}{(1+z)} t_{\rm var}$, where $\delta\simeq 10$ {\citep{Rait13}} is the Doppler factor, ultimately resulting in a lower magnetic field limit, $B$. For our purposes, for optically emitting electrons, this can be expressed { as $B \simeq 1.6 (\frac{1+z}{\delta_{\mathrm 10}})^{1/3} (\lambda_{\mathrm 6000})^{1/3} (t_{\mathrm var}, [h])^{-2/3} \mathrm ~\rm{G}$ \citep{Weav19, Fan21}, where $\delta_{\mathrm 10}=\delta/10$ and $\lambda_{\mathrm 6000}=\lambda/6000${\AA} }. If the sharpest, relatively long-lasting drop in brightness (on the night of 28 November 2024, Figure \ref{f2}) can be attributed to synchrotron losses, then $t_{\rm var}$ can be estimated to be about 20 Ks ($\sim$5 h), which leads to a lower limit of the magnetic field $B \simeq 1$ G. An even sharper increase can be seen during the previous night (27 November 2024, Figure \ref{f2}), where a $\sim$0.2 mag increase was observed for $\sim$15 min. This sharp rise cannot be attributed to energy losses, but it can be used to set an upper limit on the emitting region size, taking into account that $R\leq\frac{\delta}{1+z}ct_{\rm var}$. Thus, the size can be estimated to be $R\leq 10^{15}$ cm, provided that $\delta\sim 10$, which is a more compact emitting region than what is usually assumed (see also \citep{Maj25}).

\subsection{Inter-Band Time Delays}

Finding time delays between optical bands is a strong clue that energy losses play a primary role in generating variability. However, this and other previous studies \citep{Zhai12, Bha18, Fang22, Bach23} failed to find optical time lags or at least failed to demonstrate consistent behavior in that respect for this object. In this study, we found no convincing evidence of time lags in any of the well-sampled and rapidly variable light curves. The lags are apparently found only in datasets, where the time resolution is comparable to the measured lag or the overall variability is rather insignificant.

\subsection{Color Changes}

Color trends, and more specifically, bluer-when-brighter (BWB) trends, have been reported on some occasions on an intra-night time-scale \citep{Kali22, Imaz23, Cha23} and imply the presence of energy-density evolution (acceleration/losses) of the emitting particles as a part of the variability mechanism. Color changes will naturally occur, considering that the synchrotron losses are $\dot{\gamma}\propto-\gamma^{2}$ \citep{Ryb86}, where $\gamma$ is the individual electron’s Lorentz factor, meaning that the characteristic cooling time will scale as $\tau\propto1/\gamma$, so higher-energy emissions will evolve faster. BWB behavior can also be confirmed on long time-scales (Figure \ref{f4}), which is actually typical for this and similar BL Lac-type objects \citep{Gaur12, Bha18}.

\subsection{Emerging Picture}

Our findings support an emerging picture to explain the most notable features of the optical emissions of blazars like BL Lacertae (see also \citep{Rai23}). A curved jet modulates the change in the Doppler factor, causing high-activity states to occur, when the bulk of the emitting filaments happen to be closely aligned with the line of sight. For the distinctive \textit{{structures}} of the short-term light curves to be produced, there can be neither too many nor too few filaments. Their fast-evolving emission (in terms of both flux and polarization) is perhaps due to the passage of the emitting filaments through a standing shock while moving down the jet; alternatively, they may be overrun by a fast-moving shock. The shock locally orders the magnetic field and is able to accelerate charged particles, which can cause both rapid flux and polarization changes. The \textit{multiplicative} or avalanche-type mechanism that triggers occasional episodes of violent activity (i.e., a violent response to a small disturbance) remains to be understood \citep{Asch14, Bach25, Maj25}.

\section{Conclusions}

In this work, we analyzed optical data obtained during a recent (2024) high-activity episode of the blazar BL Lacertae. We focused on photometric and polarimetric variations on the shortest (intra-night) time-scales. We were able to compare our results with those from a previous BL Lacertae activity episode (2020). On intra-night time-scales, we detected violent flux changes with no firmly detectable inter-band lags and some polarization changes. Bluer-when-brighter behavior was evident on all scales, but no significant flux variability in flux--polarization correlations could be found. These results are comparable with the ones obtained in the previous campaign and are discussed in the context of an emerging framework to explain the violent variations in the blazar's non-thermal emissions in the optical region.

\vspace{6pt}
\authorcontributions{R.B. wrote most of the text and performed many of the observations and the analyses. M.M., A.S., and A.K. contributed to the observations. All authors suggested many changes to improve the text. All authors have read and agreed to the published version of \mbox{the manuscript.}}

\funding{This research was partially supported by the Bulgarian National Science Fund of the Ministry of Education and Science under the grants KP-06-H68/4 (2022) and KP-06-H88/4 (2024). The Skinakas Observatory is a collaborative project of the University of Crete, the Foundation for Research and Technology--Hellas, and the Max-Planck-Institut f\"ur Extraterrestrische Physik.}

\dataavailability{All published data are available upon request; however, conditions concerning their usage may apply.}

\conflictsofinterest{The authors declare no conflicts of interest. The funders had no role in the design of the study, in the collection, analyses, or interpretation of data, in the writing of the manuscript, or in the decision to publish the {results.}
}

\begin{adjustwidth}{-\extralength}{0cm}

\reftitle{References}

\PublishersNote{}
\end{adjustwidth}

\end{document}